\documentclass[final]{IEEEtran}
\usepackage{geometry}
\newgeometry{left=1.5cm,right=1.50cm,bottom=1.50cm,top=1.50cm}%
\usepackage{cite}
\usepackage{threeparttable}
\usepackage{graphicx}
\usepackage{picinpar}
\usepackage[cmex10]{amsmath}
\usepackage{amsmath,amsfonts,amssymb}
\usepackage{subfigure}
\usepackage{algorithm}
\usepackage{algorithmic}
\usepackage{stfloats}
\usepackage{bm}

\usepackage{amsthm}

\begin{document}
\newtheorem{lemma}{Lemma}
\newtheorem{corol}{Corollary}
\newtheorem{theorem}{Theorem}
\newtheorem{proposition}{Proposition}
\newtheorem{problem}{Problem}
\newtheorem{definition}{Definition}
\newcommand{\e}{\begin{equation}}
\newcommand{\ee}{\end{equation}}
\newcommand{\eqn}{\begin{eqnarray}}
\newcommand{\eeqn}{\end{eqnarray}}

\title{\LARGE Channel Estimation for Millimeter-Wave Massive MIMO with Hybrid Precoding over Frequency-Selective Fading Channels}

\author{Zhen Gao, Linglong Dai, Chen Hu, and Zhaocheng Wang
\thanks{This work was supported by the National Natural Science Foundation of China (Grant Nos. 61271266, 61571270, and 61302097),  the Beijing Natural Science Foundation (Grant No. 4142027), and the Foundation of Shenzhen government. L. Dai is the corresponding author.}
\thanks{Z. Gao, L. Dai, C. Hu, and Z. Wang are with Tsinghua National Laboratory for
 Information Science and Technology (TNList), Department of Electronic Engineering,
 Tsinghua University, Beijing 100084, China (E-mails: gao-z11@mails.tsinghua.edu.cn; \{daill,zcwang\}@mail.tsinghua.edu.cn).}\\
 \vspace*{-6.6mm}
}

\maketitle
\begin{abstract}
 Channel estimation for millimeter-wave (mmWave) massive MIMO with hybrid precoding is challenging
 , since the number of radio frequency (RF) chains is usually much smaller than that of antennas.
 To date, several channel estimation schemes have been proposed for mmWave massive MIMO over narrow-band channels, while practical mmWave channels exhibit the frequency-selective fading (FSF). 
 To this end, this letter proposes a multi-user uplink channel estimation scheme for mmWave massive MIMO
 over FSF channels. 
 Specifically, by exploiting the angle-domain structured sparsity of mmWave FSF channels, a distributed compressive sensing (DCS)-based channel estimation scheme is proposed. 
 Moreover, by using the grid matching pursuit strategy with adaptive measurement matrix, the proposed algorithm can solve the power leakage problem caused by the continuous angles of arrival or departure (AoA/AoD).
Simulation results verify that the good performance of the proposed solution. 
\end{abstract}
\vspace*{-2mm}

\IEEEpeerreviewmaketitle
\vspace*{-4mm}
\section{Introduction}\label{S1}


Millimeter-wave (mmWave) massive MIMO has been considered as a promising key technique for 5G since it can offer gigabit-per-second data rates~\cite{dong}.
For mmWave massive MIMO with the phase shifter network based hybrid precoding \cite{multi_user_precoding,myWC,dong} or electromagnetic lens based beamspace MIMO \cite{Brady,Zeng}, the number of radio frequency (RF) chains is usually much smaller than that of antennas for reduced hardware cost and power consumption.  
However, such architectures will lead to the challenging channel estimation due to only a limited number of RF chains but hundreds of antennas \cite{Han_ref}.

To date, several channel estimation schemes \cite{SED,Hea_JSTSP,Han_ref} have been proposed for mmWave massive MIMO with hybrid precoding. Specifically,
the reference signal was designed in \cite{Han_ref} for the estimation of angles of arrival or departure (AoA/AoD), but it assumes the discrete AoA/AoD. 
\cite{Hea_JSTSP} proposed an adaptive channel estimation for mmWave massive MIMO, but it is limited to single-user scenario.  
\cite{SED} can directly estimate the singular subspace with Krylov subspace method, but multiple amplify-and-forward operations between the transmitter and receiver will introduce much noise.
 More importantly, existing schemes \cite{SED,Hea_JSTSP,Han_ref} only consider the narrow-band flat fading channel model, while the practical mmWave channels exhibit the broad-band frequency-selective fading (FSF) due to the very large bandwidth and different delays of multipath \cite{FSF}.

 To this end, this letter proposes a multi-user uplink channel estimation scheme for mmWave massive MIMO systems, where the broad-band FSF channel is converted to multiple parallel narrow-band flat fading channels when OFDM is considered. Specifically, the mmWave channels exhibit
 the obviously angle-domain sparsity due to the much higher path loss for non-line-of-sight (NLOS) paths than that for  line-of-sight (LOS) paths~\cite{Hea_JSTSP}. Moreover, this sparsity is almost unchanged within the system bandwidth according to our derivation. By exploiting such angle-domain structured sparsity of
mmWave FSF channels, we propose a distributed
compressive sensing (DCS)-based channel estimation scheme, where 
both the transmit pilot signal and receive channel estimation algorithm are elaborated under the DCS theory for improved performance.
By contrast, conventional scheme in \cite{Hea_JSTSP} fails to leverage the structured sparsity of channels. Moreover, by using the grid matching pursuit strategy with adaptive measurement matrix,
the proposed algorithm can solve the power leakage problem caused by the continuous AoA/AoD.
Simulation results verify the good performance of the proposed scheme. 
 To the best of our knowledge, this is the first paper to investigate the FSF channel estimation for mmWave massive MIMO systems.

\emph{Notation}: the boldface lower and upper-case symbols denote column vectors and matrices, respectively. The Moore-Penrose inversion, transpose, conjugate transpose, integer ceiling, and expectation operators are given by $(\cdot )^{\dag}$, $(\cdot )^{\rm T}$, $(\cdot )^{*}$, $\lceil \cdot \rceil$, and ${\rm E}\{\cdot \}$, respectively. 
 $\left| \Gamma \right|$ is the
 cardinality of the set $\Gamma$. The support set of the vector $\mathbf{a}$ is denoted by
 ${\rm supp}\{\mathbf{a}\}$. 
 $ \otimes $ is
 the Kronecker product, and ${\rm{vect}}\left( {\cdot} \right)$ is the vectorization
 operation according to the columns of the matrix. $\left[ {\bf a} \right]_i$ denotes the $i$th entry of the
 vector $\mathbf{a}$, and $\left[ {\bf A} \right]_{i,j}$ denotes the $i$th-row and
 $j$th-column element of the matrix $\mathbf{A}$.

\vspace*{-3mm}
\section{System Model}\label{S2}

\begin{figure}[!tp]
\centering
\includegraphics[width=8.5cm]{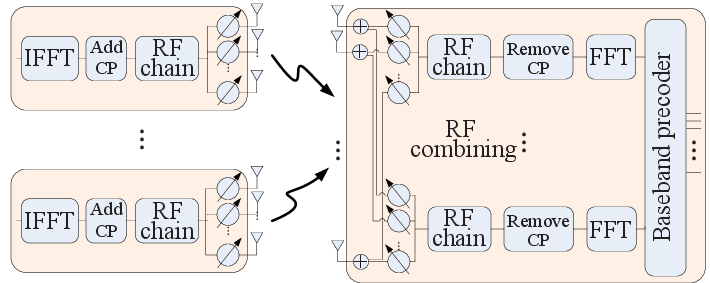}
\vspace*{-3mm}
\caption{Illustration of a multi-user broad-band mmWave massive MIMO system.} \label{fig:Spectrum}
\vspace*{-5mm}
\end{figure}

We consider a typical mmWave massive MIMO-OFDM system over FSF channels as shown in Fig. \ref{fig:Spectrum} \cite{FSF}, where the base station (BS) employs $N_a^{\rm BS}$ antennas but only $N_{\rm RF}^{\rm BS}$ RF chains with $N_a^{\rm BS}\gg N_{\rm RF}^{\rm BS}=K$ to support $K$ user equipments (UEs), and each UE has $N_a^{\rm UE}$ antennas but only $N_{\rm RF}^{\rm UE}$ RF chain with $N_a^{\rm UE}\gg N_{\rm RF}^{\rm UE}=1$. The hybrid analog-digital precoding at the BS can be used to realize the spatial multiplexing of multiple data streams with low hardware cost and energy consumption \cite{FSF}. 
Particularly, the uplink FSF channel associated with the $k$th user in the delay domain can be modeled as~\cite{FSF}
\vspace*{-1mm}
\begin{equation}\label{equ0}
{{\bf{H}}^d_k}\left( \tau  \right) = \sum\nolimits_{l = 0}^{L_k - 1} {{\bf{H}}_{l,k}^d\delta \left( {\tau  - {\tau _{l,k}}} \right)} ,\vspace*{-1mm}
\end{equation}
where $L_k$ is the number of multipath, $\tau_{l,k}$ is the delay of the $l$th path, ${{\bf{  H}}^d_{l,k}} \in  {\mathbb{C}^{{N_{a}^{\rm BS}} \times {N_a^{\rm UE}}}}$ is given by\vspace*{-1mm}
\begin{equation}
{\bf{ H}}^d_{l,k}{\rm{ = }}{\alpha _{l,k}}{{\bf{a}}_{\rm BS}}\left( {d\sin ({\theta _{l,k}})/\lambda } \right){{\bf{a}}}_{\rm UE}^*\left( d\sin ({\varphi  _{l,k}}) /\lambda \right),\label{equ1}\vspace*{-1mm}
\end{equation}
${{\alpha _{l,k}}}$ is the complex gain of the $l$th path, and ${{\theta _{l,k}}}\in [0,2\pi]$ and ${{\varphi _{l,k}}}\in [0,2\pi]$
are azimuth AoA/AoD if we consider the typical uniform linear array (ULA). For path gains, we consider Rician fading channels consisting of one LOS path (the 0th path) and $L_k-1$ NLOS paths (the $l$th path for $1\le l \le L_k-1$), where
path gains follow the mutually independent complex Gaussian distribution with zero means, and $K_{\rm factor}$ denotes the ratio
between the power of LOS path and the power of NLOS paths. 
In addition, 
\begin{equation}
\begin{small}
\!\!\!\!\begin{array}{l}
{{\bf{a}}_{{\rm{BS}}}}\left( {\frac{{d\sin ({\theta _{l,k}})}}{\lambda }} \right) = {\left[ {{e^{j2\pi {n_{{\rm{BS}}}}d\sin ({\theta _{l,k}})/\lambda }}} \right]^{\rm{T}}}_{{n_{{\rm{BS}}}} \in \left[ {0,1, \cdots ,N_a^{{\rm{BS}}}} \right]}\\
{{\bf{a}}_{{\rm{UE}}}}\left( {\frac{{d\sin ({\varphi _{l,k}})}}{\lambda }} \right) = {\left[ {{e^{j2\pi {n_{{\rm{UE}}}}d\sin ({\varphi _{l,k}})/\lambda }}} \right]^{\rm{T}}}_{{n_{{\rm{UE}}}} \in \left[ {0,1, \cdots ,N_a^{{\rm{UE}}}} \right]}
\end{array}
\end{small}
\end{equation}
are steering vectors at the BS and the $k$th user, respectively,
where $\lambda $ denotes the wavelength and $d$ is the antenna spacing.

\vspace*{-2mm}
\section{DCS-Based Channel Estimation Scheme}\label{S3}
In this section, we propose a DCS-based channel estimation scheme to jointly estimate the FSF channels. 
\vspace*{-2mm}
\subsection{Uplink Pilot Training}\label{channel estimation1}
We consider that the training sequences used for channel estimation adopt OFDM to combat the FSF channels, where lengths of cyclic prefix (CP) and  discrete Fourier transform (DFT) are ${L_{{\rm{CP}}}} > ({{\max \left\{ {{\tau _{l,k}}} \right\}_{l = 0,k = 1}^{{L_k}-1,K} - \min \left\{ {{\tau _{l,k}}} \right\}_{l = 0,k = 1}^{{L_k}-1,K}}}){{{f_s}}}$ and $P>{L_{{\rm{CP}}}}$, respectively, where $f_s$ is the sampling rate.
At the BS, after the CP removal and DFT operation, the received signal at the $p$th ($1 \le p \le P$) subcarrier
 of the $t$th OFDM symbol in the frequency domain can be expressed as
\begin{equation}\label{equ:channelmode5}  
\!\!\!\!\!\!\begin{array}{l}
{\bf{r}}_p^{(t)} = {({\bf{Z}}_{{\rm{RF}}}^{(t)}{\bf{Z}}_{{\rm{BB}},p}^{(t)})^*}\sum\nolimits_{k = 1}^K {{{\bf{ H}}_{p,k}^f}{\bf{F}}_{{\rm{RF}},k}^{\left( t \right)}{\bf{F}}_{{\rm{BB}},p,k}^{\left( t \right)}{\bf{s}}_{p,k}^{\left( t \right)}}  + {\bf{v}}_p^{(t)},
\end{array}
\vspace*{-2mm}
\end{equation}where ${\bf{r}}_p^{(t)}\in \mathbb{C}^{N_{\rm{RF}}^{\rm BS}\times 1}$ is the received signal dedicated to the $p$th pilot subcarrier in the $t$th
  OFDM symbol, ${\bf{Z}}_{{\rm{BB}},p}^{(t)}\in \mathbb{C}^{N_{\rm RF}^{\rm BS}\times N_{\rm RF}^{\rm BS}}$ is the digital combining matrix, ${\bf{Z}}_{{\rm{RF}}}^{(t)}\in \mathbb{C}^{N_{a}^{\rm BS}\times N_{\rm RF}^{\rm BS}}$ is the RF combining matrix, ${\bf{Z}}_{p}^{(t)} = {\bf{Z}}_{{\rm{RF}}}^{(t)}{\bf{Z}}_{{\rm{BB}},p}^{(t)}\in \mathbb{C}^{N_{a}^{\rm BS}\times N_{\rm RF}^{\rm BS}}$ is the composite combining matrix at the BS,
  \begin{equation}\label{equ:channelmode11}
\!\!\!\!\begin{array}{l}
{\bf{H}}_{p,k}^f = \sum\nolimits_{l = 0}^{{L_k} - 1} {{\bf{H}}_{l,k}^d{e^{ - j2\pi {f_s}{\tau _{l,k}}p/P}}} \\
= \sum\nolimits_{l = 0}^{{L_k} - 1} {{\alpha _{l,k}} {e^{ - \frac{{j2\pi {f_s}{\tau _{l,k}p}}}{P}}}{{\bf{a}}_{{\rm{BS}}}}\left( {\frac{{d\sin ({\theta _{l,k}})}}{\lambda }} \right){\bf{a}}_{{\rm{UE}}}^*\left( {\frac{{d\sin ({\varphi _{l,k}})}}{\lambda }} \right)} ,
  \end{array}
\end{equation}
denotes the frequency-domain channel matrix associated with the $p$th pilot subcarrier for the $k$th UE,
${\bf{F}}_{{\rm{RF}},k}^{\left( t \right)}\in \mathbb{C}^{N_{a}^{\rm UE}\times N_{\rm RF}^{\rm UE}}$,
  ${\bf{F}}_{{\rm{BB}},p,k}^{\left( t \right)}\in \mathbb{C}^{N_{\rm RF}^{\rm UE}\times N_{\rm RF}^{\rm UE}}$,
 ${\bf{s}}_{p,k}^{\left( t \right)}\in \mathbb{C}^{N_{\rm RF}^{\rm UE}\times 1}$ are the RF precoding matrix, digital precoding matrix, and
 transmitted training sequence for the $k$th UE, respectively, ${\bf{f}}_{p,k}^{\left( t \right)} =  {\bf{F}}_{{\rm{RF}},k}^{\left( t \right)}{\bf{F}}_{{\rm{BB}},p,k}^{\left( t \right)}{\bf{s}}_{p,k}^{\left( t \right)}\in \mathbb{C}^{N_{a}^{\rm UE}\times 1}$ is considered as the pilot signal transmitted by the $k$th user, and ${\bf{v}}_p^{(t)}$ is the additive white Gaussian noise (AWGN) at the BS. Note that RF precoding/combining is the same for all subcarriers, since the RF phase shifter network can provide constant phase shift response over a wide frequency range \cite{FSF}.

Since the path loss for NLOS paths is much larger than that for LOS paths in mmWave systems, the mmWave channels appear the obvious sparsity in the angular domain, which indicates small $L_k$ and large $K_{\rm factor}$ in mmWave systems, e.g., $L_k=4$ and $K_{\rm factor}=20$~dB \cite{dong}. Hence, we can transform the frequency-domain channel matrix ${{\bf{ H}}^f_{p,k}}$ in (\ref{equ:channelmode11}) into the sparse angle-domain channel matrix ${{\bf{ H}}^a_{p,k}}$ as \cite{Hea_JSTSP}
 \begin{equation}
{{\bf{ H}}^a_{p,k}} = {{\bf{A}}_{\rm BS}^*}{{\bf{ H}}^f_{p,k}}{\bf{A}}_{\rm UE},\label{equangular}
\end{equation}
where ${{\bf{A}}_{\rm BS}}\in \mathbb{C}^{N_a^{\rm BS}\times N_a^{\rm BS}}$ and ${{\bf{A}}_{\rm UE}}\in \mathbb{C}^{N_a^{\rm UE}\times N_a^{\rm UE}}$ are the DFT matrices by quantizing the virtual angular domain with the resolutions of $2\pi/N_a^{{\rm BS}}$ at the BS and $2\pi/N_a^{{\rm UE}}$ at the user, respectively. By vectorizing
${{\bf{ H}}^f_{p,k}}$, we can further obtain \vspace*{-2mm}
 \begin{equation}
{{\bf{h}}_{p,k}^f}\!\! =\!\! {\rm{vect}}\left(\! {{{\bf{H}}^f_{p,k}}}\! \right)\!\! = \! {({{\bf{A}}_{\rm UE}^*)}^{\rm{T}} \!\otimes \! {\bf{A}}_{\rm BS}} {\rm{vect}}\left( {{{{\bf{ H}}}^a_{p,k}}} \right) \!\!= \!\!{\bf{A}}{{\bf{ h}}^a_{p,k}},\label{equangular2}
\vspace*{-2mm}\end{equation}
where ${\bf{A}}={({{\bf{A}}_{\rm UE}^*)}^{\rm{T}} \otimes {\bf{A}}_{\rm BS}} $ and ${{\bf{ h}}^a_{p,k}}={\rm{vect}}\left( {{{{\bf{ H}}}^a_{p,k}}} \right) $.
 Due to the sparsity of ${{\bf{ H}}^a_{p,k}}$, only a minority of elements of ${\bf h}_{p,k}^a$ dominate the majority of the channel energy, and thus we have
\begin{equation}\label{equ:channelmode3} 
 \left|\Theta_{p,k}\right| = \left| {\rm supp}\left\{ {\bf h}_{p,k}^a \right\} \right| = S_k \ll N_a^{\rm BS} N_a^{\rm UE} ,
\end{equation}
 where $\Theta_{p,k}$ is the support set, and $S_k$ is the sparsity level in the angular domain. Note that if we consider
 the quantized AoA/AoD have the same resolutions as ${\bf{A}}_{\rm UE}$ and ${\bf{A}}_{\rm BS}$, we have $S_k=L_k$ \cite{Hea_JSTSP}.

According to (\ref{equangular})-(\ref{equ:channelmode3}), (\ref{equ:channelmode5}) can be further expressed as
\begin{align}\label{equ:channelmode51}  
\!\!\!\!\!\!\begin{array}{l}
{\bf{r}}_p^{(t)} = {({\bf{Z}}_p^{(t)})^*}\sum\nolimits_{k= 1}^K {{{\bf{A}}_{\rm BS}}{\bf{ H}}_{p,k}^a{\bf{A}}_{\rm UE}^*{\bf{f}}_{p,k}^{(t)}}  + {\bf{v}}_p^{(t)}\\
= {({\bf{Z}}_p^{(t)})^*}{{\bf{A}}_{\rm BS}}{{{\bf{\bar H}}}^a_p}{\bf{\bar A}}_{\rm UE}^*{\bf{\bar f}}_p^{(t)} + {\bf{v}}_p^{(t)}\\
 = \left( {{{\left( {{\bf{\bar A}}_{\rm UE}^*{\bf{\bar f}}_p^{(t)}} \right)}^{\rm{T}}} \otimes {{({\bf{Z}}_p^{(t)})}^*}{{\bf{A}}_{\rm BS}}} \right){\rm{vect}}\left( {{{{\bf{\bar H}}}^a_p}} \right) + {\bf{v}}_p^{(t)}\\
 = {{\bf{\Psi  }}_p^{\left( t \right)}}{\bf{\bar h}}_p^{a} + {\bf{v}}_p^{(t)},
 \end{array}\vspace*{-2mm}
\end{align}
where  \vspace*{-2mm}
\begin{equation}
\begin{small}
\!\begin{array}{l}
{\bf{\bar H}}_p^a = \left[ {{\bf{H}}_{p,1}^a,{\bf{H}}_{p,2}^a, \cdots ,{\bf{H}}_{p,K}^a} \right] \in \mathbb{C}{^{N_a^{{\rm{BS}}} \times KN_a^{{\rm{UE}}}}},\\
{\bf{\bar A}}_{{\rm{UE}}}^* = {\rm{diag}}\left\{ {{\bf{A}}_{{\rm{UE}}}^*,{\bf{A}}_{{\rm{UE}}}^*, \cdots ,{\bf{A}}_{{\rm{UE}}}^*} \right\} \in\mathbb{C} {^{KN_a^{{\rm{UE}}} \times KN_a^{{\rm{UE}}}}},\\
{\bf{\bar f}}_p^{\left( t \right)} = {[{({\bf{f}}_{p,1}^{(t)})^{\rm{T}}},{({\bf{f}}_{p,2}^{(t)})^{\rm{T}}}, \cdots ,{({\bf{f}}_{p,K}^{(t)})^{\rm{T}}}]^{\rm{T}}} \in \mathbb{C}{^{KN_a^{{\rm{UE}}} \times 1}},\\
{\bf{\bar h}}_p^a = {\rm{vect}}\left( {{\bf{\bar H}}_p^a} \right) \in\mathbb{C} {^{KN_a^{{\rm{BS}}}N_a^{{\rm{UE}}} \times 1}},\\
{\bf{\Psi }}_p^{\left( t \right)} = {\left( {{\bf{\bar A}}_{{\rm{UE}}}^*{\bf{\bar f}}_p^{(t)}} \right)^{\rm{T}}} \otimes {({\bf{Z}}_p^{(t)})^*}{{\bf{A}}_{{\rm{BS}}}} \in \mathbb{C}{^{N_{{\rm{RF}}}^{{\rm{BS}}} \times KN_a^{{\rm{BS}}}N_a^{{\rm{UE}}}}}.
\end{array}\vspace*{-2mm}
\end{small}
\end{equation}

 Furthermore, we consider the mmWave channels remain unchanged in $G$ successive OFDM symbols within the channel coherence time \cite{Hea_JSTSP}. By jointly using the received pilot signals in $G$ successive OFDM symbols, we can obtain\vspace*{-2mm}
\begin{align}\label{equ:joint_process2} 
{\bf{\tilde r}}_p^{} = {\bf{\tilde \Psi }}_p^{}\bar {\bf{h}} _p^a + {\bf{\tilde v}}_p^{},
\end{align}
where ${\bf{\tilde r}}_p^{} \!\!= \!\!{[ {{{( {{\bf{r}}_p^{(1)}} )^{\rm{T}}}},{{( {{\bf{r}}_p^{(2)}} )^{\rm{T}}}}, \cdots ,{{( {{\bf{r}}_p^{(G)}} )^{\rm{T}}}}} ]^{\rm{T}}}\!\!\in \mathbb{C}^{GN_{\rm RF}^{\rm BS}\times 1}$ is the aggregate received signal,
${{\bf{\tilde \Psi  }}_p^{}} \!\!\!\!= \!\!\!\!{[ {{{( {{{\bf{\Psi  }}_p^{(1)}}} )^{\rm{T}}}},{{( {{{\bf{\Psi  }}_p^{(2)}}} )^{\rm{T}}}}, \cdots ,{{( {{{\bf{ \Psi }}_p^{(G)}}} )^{\rm{T}}}}} ]^{\rm{T}}}\!\!\in\!\! \mathbb{C}^{GN_{\rm RF}^{\rm BS} \times KN_{a}^{\rm BS}N_{a}^{\rm UE}}$ is the aggregate measurement matrix,
and ${\bf{\tilde v}}_p^{} \!\!= \!\!{[ {{{( {{\bf{v}}_p^{(1)}} )^{\rm{T}}}},{{( {{\bf{v}}_p^{(2)}} )^{\rm{T}}}}, \cdots ,{{( {{\bf{v}}_p^{(G)}} )^{\rm{T}}}}} ]^{\rm{T}}}$ is aggregate AWGN. The system's SNR
 can be defined as $\mbox{SNR}\!\!=\!\!{\rm E}\Big\{\Big\|{\bf \tilde \Psi }_p^{}
 \bar{\bf h}_p^{a }\Big\|_2^2\Big\} \Big/ {\rm E}\Big\{\Big\| {\bf \tilde v}_p^{}\Big\|_2^2\Big\}$ according to (\ref{equ:joint_process2}).
 \vspace*{-4mm}
\subsection{DCS-Based Channel Estimation}\label{channel estimation}
 To accurately estimate channels 
 from (\ref{equ:joint_process2}), $G$ in conventional algorithms,
 such as the minimum mean square error (MMSE) algorithm, is heavily dependent on the
 dimension of $\bar {\bf{h}} _p^a$, i.e., $KN_{a}^{\rm UE}N_{a}^{\rm BS}$. Usually, $GN_{\rm RF}^{\rm BS}\ge KN_{a}^{\rm UE}N_{a}^{\rm BS}$ is required,
 which leads $G$ to be much larger than the channel coherence time
 ~\cite{Hea_JSTSP}. Fortunately, the sparsity of mmWave massive MIMO channels 
 motivates us to leverage the CS theory to estimate channels with much reduced pilot overhead. Moreover, according to (\ref{equ:channelmode11}), it can
 be observed that $\{ {{\bf{H}}_{p,k}^f} \}_{p = 1}^P$ share the same AoA/AoD, and thus $\{{\bf h}^a_{p,k}\}_{p=1}^{P}$ obtained after (\ref{equangular}) and (\ref{equangular2}) have the structured sparsity within the system bandwidth, i.e.,
 \begin{equation}\label{eq5}
{\rm supp}\left\{{\bf h}^a_{1,k}\right\}\! = \!{\rm supp}\left\{{\bf h}_{2,k}^a\right\}\!
 = \!\cdots \!= \!{\rm supp}\left\{{\bf h}^a_{P,k}\right\}\! = \!\Theta_k.
\end{equation}
 Specifically, given (\ref{equ:joint_process2}) and the sparse constraints of (\ref{equ:channelmode3}) and (\ref{eq5}), the channels can be estimated with standard DCS tool. 
 However, due to the continuous AoA/AoD and the limited angle-domain resolution of ${\bf{A}}_{\rm BS}$ and ${\bf{A}}_{\rm UE}$, the sparsity of $\bar {\bf{h}} _p^a$ may be impaired due to the power leakage problem \cite{Hea_JSTSP}, which will result in the poor channel estimation performance. 

To this end,
  we propose a distributed grid matching pursuit (DGMP) algorithm as listed in \textbf{Algorithm 1} including
  outer loop and inner loop.
  In each iteration of outer loop (steps 2.1-2.3 and 2.19-2.21),
  according to correlation operation (step 2.1), the UE index $\tilde k$ (step 2.2) and adaptive measurement matrix ${\bf {\bar\Upsilon}}_p$ (step 2.3) associated with the most possible path are acquired and input to inner loop; according to the output of inner loop, the $\tilde k$th UE's transmit/receive steering vectors are acquired (steps 2.19-2.20), and $|\cal{K}|$ UEs' LOS path gains and residue ${\bf b}_p$ are updated (step 2.21). The iteration of outer loop stops when AoA/AoD and path gains of all $K$ UEs' LOS paths are estimated.
  For inner loop (steps 2.4-2.18), the AoA/AoD estimation associated with the $\tilde k$th UE's
  LOS path is improved with the grid matching strategy. Specifically,
  according to the inputs $\tilde k$ and ${\bf {\bar\Upsilon}}_p$ from outer loop,
  AoA/AoD indices $n^{\rm BS}$ and $n^{\rm UE}$ of the most possible path are acquired (step 2.6),
  and the corresponding correlation value is recorded as $\beta$ (step 2.5);
  we construct the local over-complete
  measurement matrix ${\bf{\tilde \Upsilon }}_p^{}$ (steps 2.7-2.11),
  where the local resolution of AoA associated with the index $n^{\rm BS}$
  and AoD associated with the index $n^{\rm UE}$
  is increased by $(2J-1)$ times;
  according to correlation operation (step 2.12),
  finer AoA/AoD indices $m^{\rm BS}$ and $m^{\rm UE}$ can be acquired (step 2.13);
  finally, ${\bf{\bar\Upsilon }}_p^{}$ is adaptively updated,
  where the grid of AoA/AoD candidates is adjusted according to $m^{\rm BS}$ and $m^{\rm UE}$ (step 2.14-2.18). The iteration of inner loop stops when $\left| {{\beta _{{\rm{last}}}} - \beta } \right| < \varepsilon $.






With the joint process of ${{\bf{\tilde \Psi }}_p^{}}$ and ${\bf{b}}_p$ for $1\le p \le P$, the DGMP algorithm exploits the structured sparsity for improved performance, which can be found in steps 2.1, 2.4, and 2.12. Moreover, the adaptive measurement matrix ${\bf {\bar\Upsilon}}_p$ with grid matching pursuit strategy can achieve high resolution estimation of AoA/AoD. Additionally, the near-LOS mmWave channel property is exploited, where only $K$ USs' LOS paths are estimated.
Compared to the adaptive CS algorithm \cite{Hea_JSTSP} estimating single
 sparse narrow-band channel from single received signal, the proposed DGMP algorithm
  jointly estimates multiple sparse subchannels from multiple received signals, whereby the
angle-domain structured sparsity of mmWave FSF channels is exploited for improved performance.
Moreover, the grid matching pursuit strategy ({steps 2.4-2.18}) with adaptive ${\bf{\bar\Upsilon }}_p^{}$ can
solve the problem of power leakage caused by the continuous AoA/AoD, which is different from the classical DCS algorithms \cite{STR_CS}.
\vspace*{-1mm}

\begin{algorithm}[tp]
{\small
\renewcommand{\algorithmicrequire}{\textbf{Input:}}
\renewcommand\algorithmicensure {\textbf{Output:} }
\caption{Proposed DGMP Algorithm.}
\label{alg:Framwork} 
\begin{algorithmic}[1]
\REQUIRE
Received signals ${\bf{\tilde r}}_p^{}$ and sensing matrices ${\bf{\tilde \Psi }}_p^{}$
 in (\ref{equ:joint_process2}) $\forall p$, AoA/AoD resolution factor $J$, and error threshold $\varepsilon $.
\ENSURE
The steering vector estimation of $k$th UE's LOS path ${\bf{\hat a}}_{{\rm{BS}}}^{k,{\rm{LOS}}}$ and ${\bf{\hat a}}_{{\rm{UE}}}^{k,{\rm{LOS}}}$, and the estimation set of path gains $\hat {\bm{\alpha }} \in \mathbb{C}^{1\times K}$, where ${\left[ {\hat {\bm{\alpha }}} \right]_k}$ denotes the gain estimate of $k$th UE's LOS path. \\

${\kern -7pt}$$ \bullet $ \textbf{Step 1} (\emph{Initialization}) The residue ${\bf b}_p \!=\!{\bf \tilde r}_p^{}$, the iteration index $k= 1$, {\scriptsize${\left[ {{\bf{\tilde \Psi }}_p^{}} \right]_{:,j}} = {\left[ {{\bf{\tilde \Psi }}_p^{}} \right]_{:,j}}/{\left\| {{{\left[ {{\bf{\tilde \Psi }}_p^{}} \right]}_{:,j}}} \right\|_2}$} for $1\le j \le KN_a^{\rm UE} N_a^{\rm BS}$, $\forall p$, and the matrix ${\bf{\Xi }}_p$ and set $\cal{K}$ are set to be
empty.
\\
${\kern -7pt}$$ \bullet $ \textbf{Step 2} (\emph{Estimate steering vectors and gains of $K$ UEs' LOS paths}) \\
 $ \textbf{for}~k\le K ~\textbf{do} $   \\
$ \text{1.}$~~~{\scriptsize${\rho } \! = \!\arg \max\limits_{\widetilde{\rho}} \left\{ \! \sum\nolimits_{p=1}^{P} \!
 \left\| \left[ {{{\left( {{\bf{\tilde \Psi  }}_p^{}} \right)}^*}{\bf{b}}_p^{}} \right]_{\widetilde{\rho}} \right\|_2^2, \left\lceil {{\widetilde{\rho}} /(N_a^{{\rm{UE}}}N_a^{{\rm{BS}}})} \right\rceil  \notin {\cal K}  \right\}$};\\
$ \text{2.}$~~~{\scriptsize$\tilde k \!\!=\!\!\left\lceil \!{{\rho }/({{N_a^{{\rm{UE}}}N_a^{{\rm{BS}}}}})} \right\rceil$, ${\cal{K}}={\cal{K}} \cup \tilde k$}; \\
$ \text{3.}$~~~{\scriptsize${\bf {\bar  \Upsilon}}_p={\left[ {{\bf{\tilde \Psi }}_p^{}} \right]_{( {\tilde k - 1} ){N_a^{{\rm{BS}}}}{N_a^{{\rm{ UE}}}} + 1:\tilde k{N_a^{{\rm{BS}}}}{N_a^{{\rm{UE}}}},:}} $};\\
 $ \text{~~}$    $ \textbf{repeat} $ 
 \\
   $ \text{4.}$~~~{\scriptsize${\rho } \! = \!\arg \max\limits_{\widetilde{\rho}} \left\{ \! \sum\nolimits_{p=1}^{P} \!
 \left\| \left[ {{{\left( {\bf {\bar\Upsilon}}_p \right)}^*}{\bf{b}}_p^{}} \right]_{\widetilde{\rho}} \right\|_2^2  \right\}$};\\
   $ \text{5.}$~~~{\scriptsize${\beta _{{\rm{last}}}} = \beta $,~$\beta  = \sum\nolimits_{p = 1}^P  \left\| {{{\left[ {{{\left( {{{\bf{\bar\Upsilon }}_p}} \right)}^*}{\bf{b}}_p^{}} \right]}_\rho }} \right\|_2^2$};\\
  $ \text{6.}$~~~{\scriptsize$n^{{\rm{UE}}} = \left\lceil {\rho /N_a^{{\rm{BS}}}} \right\rceil $, $n^{{\rm{BS}}} = \rho  - (n^{{\rm{UE}}} - 1)N_a^{{\rm{BS}}}$};\\
   $ \text{7.}$~~~{\scriptsize${\bf{\tilde A}}_{{{\rm{UE}}}}^{}\!\! = \!\!{\left[ {{{\bf{a}}_{{\rm{UE}}}}\left(( {n^{{\rm{UE}}} + \frac{{{j_{{\rm{UE}}}}}}{{2J}})/N_a^{{\rm{UE}}}} \right)} \right]_{{j_{{\rm{UE}}}} \in \left[ { - J, - J + 1, \cdots ,J} \right]}}$}; \\
    $ \text{8.}$ ~~{\scriptsize${\bf{\tilde A}}_{{{\rm{BS}}}}^{} \!\! = \! \!{\left[ {{{\bf{a}}_{{\rm{BS}}}}\left(( {n^{{\rm{BS}}} + \frac{{{j_{{\rm{BS}}}}}}{{2J}})/N_a^{{\rm{BS}}}} \right)} \right]_{{j_{{\rm{BS}}}} \in \left[ { - J, - J + 1, \cdots ,J} \right]}}$};\\
    $ \text{9.}$~~~{\scriptsize${{\bf{\tilde \Upsilon  }}_p^{\left(t \right)}} = {\left( {{\bf{\tilde A}}_{{{\rm{UE}}}}^*{\bf{f}}_{p,\tilde k}^{(t)}} \right)^{\rm{T}}} \otimes {({\bf{Z}}_p^{(t)})^*}{{\bf{\tilde A}}_{{{\rm{BS}}}}}$};\\
    $ \text{10.}$~~{\scriptsize${\bf{\tilde \Upsilon }}_p^{} = {[{({\bf{\tilde \Upsilon }}_p^{\left( {1} \right)})^{\rm{T}}},{({\bf{\tilde \Upsilon }}_p^{\left( {2} \right)})^{\rm{T}}}, \cdots ,{({\bf{\tilde \Upsilon }}_p^{\left( {G} \right)})^{\rm{T}}}]^{\rm{T}}}$}; \\
    $ \text{11.}$~~{\scriptsize${\left[ {{\bf{\tilde \Upsilon }}_p^{}} \right]_{:,j}} = {\left[ {{\bf{\tilde \Upsilon }}_p^{}} \right]_{:,j}}/{\left\| {{{\left[ {{\bf{\tilde \Upsilon }}_p^{}} \right]}_{:,j}}} \right\|_2}$, $1\le j \le (2J-1)^2$}, $\forall p$; 
    \\
    $ \text{12.}$~~{\scriptsize${\eta } \! = \!\arg \max\limits_{\widetilde{\eta}} \left\{ \! \sum\nolimits_{p=1}^{P} \!
 \left\| \left[ {{{( {{\bf{\tilde \Upsilon  }}_p^{}} )}^*}{\bf{b}}_p^{}} \right]_{\widetilde{\eta }} \right\|_2^2  \right\}$; 
 }\\
 $ \text{13.}$~~{\scriptsize$m^{{\rm{UE}}} = \left\lceil {\eta /(2J-1)} \right\rceil $, $m^{{\rm{BS}}} = \eta  - (m^{{\rm{UE}}} - 1)(2J-1)$};\\
 $ \text{14.}$~~~{\scriptsize${\bf{\tilde A}}_{{{\rm{UE}}}}^{}\!\! = \!\!{\left[ {{{\bf{a}}_{{\rm{UE}}}}\left(( {n^{{\rm{UE}}} + \frac{{{ -J+m^{\rm UE}-1  }}}{{2J}})/N_a^{{\rm{UE}}}} \right)} \right]_{{n^{{\rm{UE}}}} \in \left[ { 0,  1, \cdots ,N_{\rm UE}-1} \right]}}$}; \\
    $ \text{15.}$ ~~{\scriptsize${\bf{\tilde A}}_{{{\rm{BS}}}}^{} \!\! = \! \!{\left[ {{{\bf{a}}_{{\rm{BS}}}}\left(( {n^{{\rm{BS}}} + \frac{ -J+m^{\rm BS}-1 }{{2J}})/N_a^{{\rm{BS}}}} \right)} \right]_{{n^{{\rm{BS}}}} \in \left[ { 0,  1, \cdots ,N_{\rm BS}-1} \right]}}$};\\
    $ \text{16.}$~~{\scriptsize${{\bf{\Upsilon  }}_p^{\left(t \right)}} = {\left( {{\bf{\tilde A}}_{{{\rm{UE}}}}^*{\bf{f}}_{p,\tilde k}^{(t)}} \right)^{\rm{T}}} \otimes {({\bf{Z}}_p^{(t)})^*}{{\bf{\tilde A}}_{{{\rm{BS}}}}}$};\\
     $ \text{17.}$~~{\scriptsize${\bf{\Upsilon }}_p^{} = {[{({\bf{\Upsilon }}_p^{\left( {1} \right)})^{\rm{T}}},{({\bf{\Upsilon }}_p^{\left( {2} \right)})^{\rm{T}}}, \cdots ,{({\bf{\Upsilon }}_p^{\left( {G} \right)})^{\rm{T}}}]^{\rm{T}}}$}; 
    \\
    $ \text{18.}$~~{\scriptsize${\left[ {{\bf{\bar \Upsilon }}_p^{}} \right]_{:,j}} = {\left[ {{\bf{ \Upsilon }}_p^{}} \right]_{:,j}}/{\left\| {{{\left[ {{\bf{ \Upsilon }}_p^{}} \right]}_{:,j}}} \right\|_2}$, $1\le j \le N_a^{\rm UE}N_a^{\rm BS}$, $\forall p$}; 
    \\
$ \text{~~}$  $ \textbf{until}$ $\left| {{\beta _{{\rm{last}}}} - \beta } \right| < \varepsilon $   \\
  $ \text{19.}$$ \text{~~}$ ${\bf{\hat a}}_{{\rm{BS}}}^{{\tilde k},{\rm{LOS}}} = {\bf{a}}_{{\rm{BS}}}^{}(({n^{{\rm{BS}}}} + \frac{{ - J + {m^{{\rm{BS}}}} - 1}}{{2J}})/N_a^{{\rm{BS}}})$;\\
     $ \text{20.}$$ \text{~~}$  ${\bf{\hat a}}_{{\rm{UE}}}^{{\tilde k},{\rm{LOS}}} = {\bf{a}}_{{\rm{UE}}}^{}(({n^{{\rm{UE}}}} + \frac{{ - J + {m^{{\rm{UE}}}} - 1}}{{2J}})/N_a^{{\rm{UE}}})$;\\
     $ \text{21.}$$ \text{~~}$  ${\bf{\Xi }}_p = \left[ {{\bf{\Xi }}_p,{{\left[ {{{\bf{\Upsilon }}_p}} \right]}_{:,\eta }}} \right]$, ${\hat {\bm{\alpha }}_{\cal K}} = {({{\bf \Xi} _p})^\dag }{{{\bf{\tilde r}}}_p}$, $ {\bf{b}}_p = {\bf{\tilde r}}_p^{} - {{\bm \alpha} _{\cal K}}{\bf{\Xi }}_p$;\\
       $ \textbf{end for}$   \\

\end{algorithmic}}
\end{algorithm}

\vspace*{-3mm}
\subsection{Pilot Design According to DCS Theory}\label{non-orthogonal}

The measurement matrices ${\bf{\tilde \Psi }}_p^{}$, $\forall p$ in
 (\ref{equ:joint_process2}) are very important for guaranteeing the reliable channel estimation. Usually, we have $GN_{\rm RF}^{\rm BS}\ll KN_{a}^{\rm UE}N_{a}^{\rm BS}$. Since ${{\bf{\tilde \Psi  }}_p^{}} \!\!\!\!= \!\!\!\!{[ {{{( {{{\bf{\Psi  }}_p^{(1)}}} )^{\rm{T}}}},{{( {{{\bf{\Psi  }}_p^{(2)}}} )^{\rm{T}}}}, \cdots ,{{( {{{\bf{ \Psi }}_p^{(G)}}} )^{\rm{T}}}}} ]^{\rm{T}}}$,
  ${{\bf{\Psi  }}_p^{\left( t \right)}} = {( {{\bf{\bar A}}_{\rm UE}^*{\bf{\bar f}}_p^{(t)}} )^{\rm{T}}} \otimes {({\bf{Z}}_p^{(t)})^*}{{\bf{A}}_{\rm BS}}$, ${\bf{\bar A}}_{\rm UE}^* = {\rm{diag}}\left\{ {{\bf{A}}_{\rm UE}^*,{\bf{A}}_{\rm UE}^*, \cdots ,{\bf{A}}_{\rm UE}^*} \right\}$, and ${\bf{ A}}_{\rm UE}$, ${\bf{ A}}_{\rm BS}$ are determined by the geometrical
  structure of
 the antenna arrays, both $\{ {{\bf{f}}_{p,k}^{(t)}} \}_{p = 1,k = 1,t=1}^{P,K,G}$ transmitted
 by the $K$ users and  $\{ {{{\bf{Z}}_p^{(t)}}} \}_{p = 1,t=1}^{P,G}$ at the BS should be elaborated
 to guarantee the desired robust channel estimation.

 According to \cite{STR_CS}, a measurement matrix whose elements follow an independent identically distributed (i.i.d.)
 Gaussian distribution can achieve the good performance for sparse signal recovery. Furthermore, diversifying measurement matrices
 ${\bf{\tilde \Psi }}_p^{}$, $\forall p$ can further improve the recovery performance of sparse signals according to DCS theory \cite{STR_CS}. This enlightens us to
 appropriately design pilot signals for mmWave massive MIMO systems.
 Specifically, as discussed above, ${\bf{Z}}_p^{(t)} = {\bf{Z}}_{{\rm{RF}}}^{(t)}{\bf{Z}}_{{\rm{BB}},p}^{(t)}$,
   ${\bf{f}}_{p,k}^{\left( t \right)} = {\bf{F}}_{{\rm{RF}},k}^{\left( t \right)}{\bf{F}}_{{\rm{BB}},p,k}^{\left( t \right)}{\bf{s}}_{p,k}^{\left( t \right)}= {\bf{F}}_{{\rm{RF}},k}^{\left( t \right)}{\bf{\tilde s}}_{p,k}^{\left( t \right)}$ if we define ${\bf{\tilde s}}_{p,k}^{\left( t \right)}={\bf{F}}_{{\rm{BB}},p,k}^{\left( t \right)}{\bf{s}}_{p,k}^{\left( t \right)}$ ($1\le k \le K$, $1 \le t \le G$, $1\le p \le P$). Hence, we propose that each element of pilot signals is given by
     \vspace*{-2mm}
\begin{equation}
\begin{small}
\!\!\!\!\!\!\!\!\!\!\!\!\!\!\!\!\!\!\!\begin{array}{l}
{\left[ {{\bf{Z}}_{{\rm{RF}}}^{(t)}} \right]_{{i_1},{j_1}}} \!\!\!\!\!\!\!\!\!\!=\! {e^{j\phi _{{i_1},{j_1},t}^1}},1 \le {i_1} \le N_a^{{\rm{BS}}},1 \le {j_1} \le N_{{\rm{RF}}}^{{\rm{BS}}},\nonumber
\end{array}
\end{small}
\end{equation}
\begin{equation}
\begin{small}
\!\!\!\!\!\!\!\!\!\!\!\begin{array}{l}
{\left[ {{\bf{F}}_{{\rm{RF}},k}^{\left( t \right)}} \right]_{{i_2},{j_2}}}  \!\!\!\!\!\!\!\!\!\!=\! {e^{j\phi _{{i_2},{j_2},t,k}^2}},1 \le {i_2} \le N_a^{{\rm{UE}}},1 \le {j_2}\! \le\! N_{{\rm{RF}}}^{{\rm{UE}}},\nonumber
\end{array}
\end{small}
\end{equation}
\begin{equation}
\begin{small}
\!\!\!\!\!\begin{array}{l}
{\left[ {{\bf{Z}}_{{\rm{BB}},p}^{(t)}} \right]_{{i_4},{j_4}}} \!\!\!\!=\! {e^{j\phi _{{i_4},{j_4},p,t}^4}},1 \le {i_4} \le N_{{\rm{RF}}}^{{\rm{BS}}},1 \le {j_4} \le N_{{\rm{RF}}}^{{\rm{BS}}},\vspace*{-2mm}\nonumber
\end{array}
\end{small}
\end{equation}
\begin{equation}
\begin{small}
\!\!\!\!\!\!\begin{array}{l}
{\left[ {{\bf{\tilde s}}_{p,k}^{\left( t \right)}} \right]_{{i_3}}}  \!\!\!\!=\! {e^{j\phi _{{i_3},p,t,k}^3}},1 \le {i_3} \le N_{{\rm{RF}}}^{{\rm{UE}}},
\end{array}
\end{small}
\end{equation}
 where ${\phi^1_{i_1,j_1,t}}$, ${\phi^2 _{i_2,j_2,t,k}}$, ${\phi^3 _{i_3,p,t,k}}$, and ${\phi^4_{i_4,j_4,p,t}}$ follow the
 i.i.d. uniform distribution ${\cal{U}}\left[ 0, ~ 2\pi\right)$. Note that elements of RF precoding/combining matrices should meet the constant modulus property, and different subcarriers share the same RF precoding/combining. It is readily seen that
 the designed pilot signals guarantee that the elements of
 ${\bf{\tilde \Psi }}_p^{}$ obey the i.i.d. complex
 Gaussian distribution with zero mean.
Moreover, ${\bf{\tilde \Psi }}_p^{}$ with different $p$ are diversified.
 Hence, the proposed pilot signal design is optimal in terms of the joint recovery of multi-user's sparse angle-domain channels in the uplink.

\vspace*{-3mm}
\section{Simulation Results}\label{S5}

 In this section, we investigate the performance of the proposed DCS-based channel estimation. 
 In simulations, carrier frequency $f_c=30$\,GHz, $f_s=0.25$\,GHz, the maximum delay spread $\tau_{\rm max}=100$ ns, $L_{\rm CP}=\tau_{\rm max}f_s=25$, $P=32$,
 $N_a^{\rm UE}=32$,
 $N_{\rm RF}^{\rm UE}=1$, $N_{a}^{\rm BS}=128$, $N_{\rm RF}^{\rm BS}=4$, $d ={\lambda}/{2}$, $K_{\rm factor}=20$~dB, $J=10$, $\varepsilon=10^{-3} $, $K=4$,
 $L_k=4$ for $1\le k\le K$. The case with the ideal AoD/AoD known at the BS is used as the performance benchmark
 for comparison.
 The adaptive CS-based channel estimation scheme \cite{Hea_JSTSP} is also adopted for comparison.

\begin{figure}[tp!]
\vspace*{-3mm}
\begin{center}
\includegraphics[width=0.85\columnwidth, keepaspectratio]{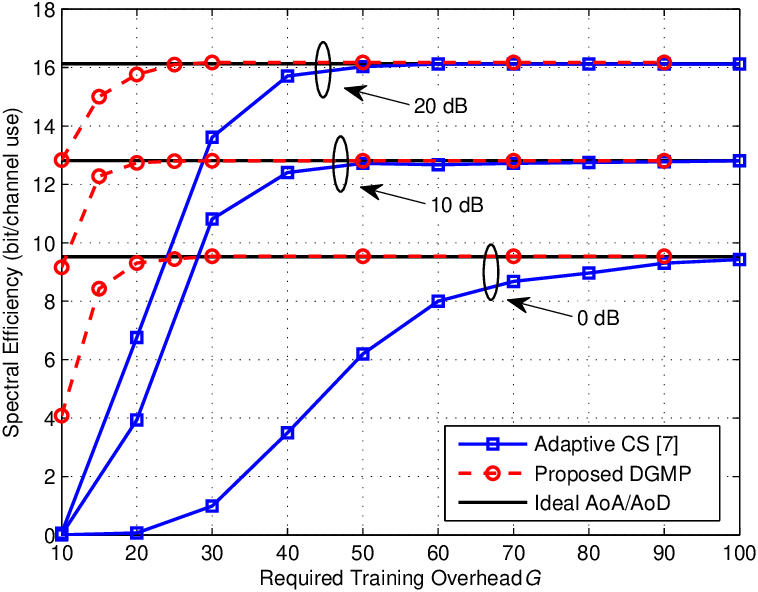}
\end{center}
\vspace*{-4mm}
\caption{Comparison of spectral efficiency performance of different channel estimation schemes against the training overhead $G$ and SNR.}
\label{fig:mse_vs_T} 
\vspace*{-5mm}
\end{figure}

Fig.~\ref{fig:mse_vs_T} investigates the downlink spectral efficiency (bit per channel use [bpcu]) by using the hybrid analog-digital precoding scheme in \cite{multi_user_precoding}, where the channels were estimated by the adaptive CS scheme \cite{Hea_JSTSP} and the proposed DGMP algorithm. The case with ideal AoA/AoD
 was adopted as the
 performance bound. From Fig.~\ref{fig:mse_vs_T}, it can be observed that the adaptive CS scheme performs poorly, since it does not
 exploit the structured sparsity of mmWave massive MIMO channels.
 In contrast, the proposed DGMP algorithm can approach the performance bound with ideal AoA/AoD when $G \ge 20$. This is because the proposed
 DCS-based channel estimation scheme can leverage the angle-domain structured sparsity of mmWave FSF channels within the system bandwidth.
By contrast, to approach the performance bound, the conventional adaptive CS algorithm requires larger $G$, e.g., $G>90$ is required at SNR = 0 dB. Hence, the proposed scheme can substantially reduce the required training overhead for FSF channel estimation compared to its counterpart.

\begin{figure}[!t]
\begin{center}
\vspace*{-3mm}
\includegraphics[width=0.85\columnwidth, keepaspectratio]{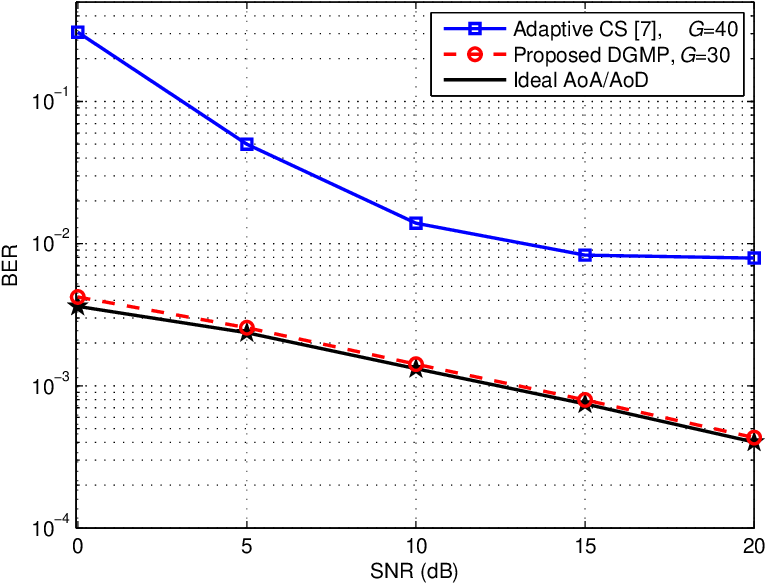}
\end{center}
\vspace*{-4mm}
\caption{BER performance comparison of different channel estimation schemes.}
\label{fig:ber_vs_snr} 
\vspace*{-6mm}
\end{figure}

 Fig.~\ref{fig:ber_vs_snr} compares the downlink bit error rate (BER) performance, where 16-QAM is used, and $G$ for adaptive CS algorithm and DGMP algorithm are 40, and 30, respectively. It can be
 observed that the proposed channel estimation scheme outperforms its counterpart with reduced training overhead,
 and its BER performance is very close to the performance bound with ideal AoA/AoD.

\vspace*{-4mm}
\section{Conclusions}\label{S6}

In this paper, we have proposed a DCS-based uplink channel estimation scheme for the
multi-user mmWave massive MIMO, which can effectively combat mmWave FSF channels. 
Specifically, we have designed an efficient pilot scheme and proposed a reliable DGMP algorithm under the framework of DCS theory, whereby the angle-domain structured sparsity of mmWave FSF channels is exploited for the
reduced training overhead.
 Moreover, by using the grid matching pursuit strategy with adaptive measurement matrix, the
proposed algorithm can effectively solve the power leakage problem.
Simulation results
 have confirmed that our scheme can accurately estimate the FSF channels in mmWave massive MIMO with much lower pilot overhead than the existing scheme.

\vspace*{-15mm}

\end{document}